\documentclass[prb,preprintnumbers,amsmath,amssymb,twocolumn]{revtex4-2}

\usepackage{graphicx}
\usepackage{dcolumn}
\usepackage{bm}

\begin{document}
\title{A comparison of the hidden order transition in URu$_2$Si$_2$ to the $\lambda$-transition in $^4$He}

\author{W. Montfrooij$^1$ and J. A. Mydosh$^2$} \affiliation{$^1$Department of Physics and Astronomy, University of Missouri, Columbia, Missouri 65211, USA\\$^2$Institute Lorentz, Leiden University, NL-2300 RA Leiden, the Netherlands}
\begin{abstract}
{The low-temperature states of ambient URu$_2$Si$_2$ and superfluid $^4$He are both characterized by momentum-dependent energy gaps between the ground and excited states. This behavior weakly persists even above the transition temperatures but becomes over-damped (ungapped) because of the number of excitations present at elevated temperature. We show that akin to the normal fluid to superfluid transition in $^4$He, the hidden-order (HO) transition in URu$_2$Si$_2$ can be understood by a change of the ungapped excitations to the gapped, elementary excitations (EE) of the unknown ordered state. These under-damped EEs reflect the basic character and order parameters of the different phase transitions. This view accounts for the full amount of entropy released in these transitions, the jumps in the resistivity and thermal conductivity directly below the transition, as well as the reduction of the Fermi surface. We argue that the behavior in the HO phase is that of a gas of weakly interacting excitations from charge density wave or crystal field states in a similar manner to that of the phonon-roton excitations of the superfluid $^4$He phase. We discuss the influence of applying pressure and magnetic fields within this scenario and the role of the small moment antiferromagnetic clustering in the hidden order phase.}
\end{abstract}
\maketitle
\section{introduction}
The hidden order transition in URu$_2$Si$_2$ that occurs at $T_{HO}$= 17.5 K under ambient presence in the absence of magnetic fields, has been studied extensively \cite{mydosh11,mydosh20} but proven very difficult to pin down. The transition itself is marked by a sharp $\lambda$-shaped increase \cite{palstra} in specific heat (on cooling), an increase in resistivity and thermal conductivity \cite{behnia,sharma} just below the transition, and a notable reduction \cite{elgazzar} of the Fermi surface. In addition, anti-ferromagnetic (AF) order appears at the transition, but the uranium moment associated with this ordering is very small \cite{broholm} and the ordering itself is not truly long range. In addition, the system undergoes a superconducting transition at T= 1.5 K. Applying hydrostatic pressure \cite{villaume} increases both $T_{HO}$ and the size of the ordered AF-moment, resulting in a first-order transition at a critical pressure of $\sim$0.5 GPa.\cite{hassinger} The application of modest magnetic fields lowers $T_{HO}$ without affecting \cite{dijk97} the amount of entropy associated with the transition.\\

Neutron scattering experiments on URu$_2$Si$_2$ \cite{broholm} have revealed that the excitations present in the system are (most likely) propagating crystal field excitations, also called van Vleck excitons \cite{mineev}, between two singlet levels. These excitations are gapped throughout the entire Brillouin zone at low temperature, but their damping increases on raising the temperature and they become strongly damped and over-damped (at select wave vectors) \cite{bourdarot10,niklowitz} at $T_{HO}$. The low-T dispersion shows a minimum at finite energy, well separated from the zero-energy intensity associated with the small moment AF-order, at the AF-wave vector $Q_0$=(1,0,0) and a higher minimum at the incommensurate vector $Q_1$=(0.6,0,0). The excitations at $Q_0$ and $Q_1$ become over-damped \cite{niklowitz} on raising the temperature through the transition, but can still easily be identified \cite{wiebe07}. In addition, ARPES and STM measurements \cite{aynajian,schmidt,mydosh11} have yielded a trove of experimental data, but despite an impressive amount of theoretical work, the order parameter associated with the hidden order transition has remained elusive. Lastly, Butch {\it et al.} \cite{butch} have presented evidence that the magnetic excitations above and below $T_{HO}$ have the same symmetry (body-centered tetragonal) suggesting an order parameter that does not break spatial symmetry.\\

\begin{figure}[t]
	\includegraphics*[viewport=90 80 625 420,width=85mm,clip]{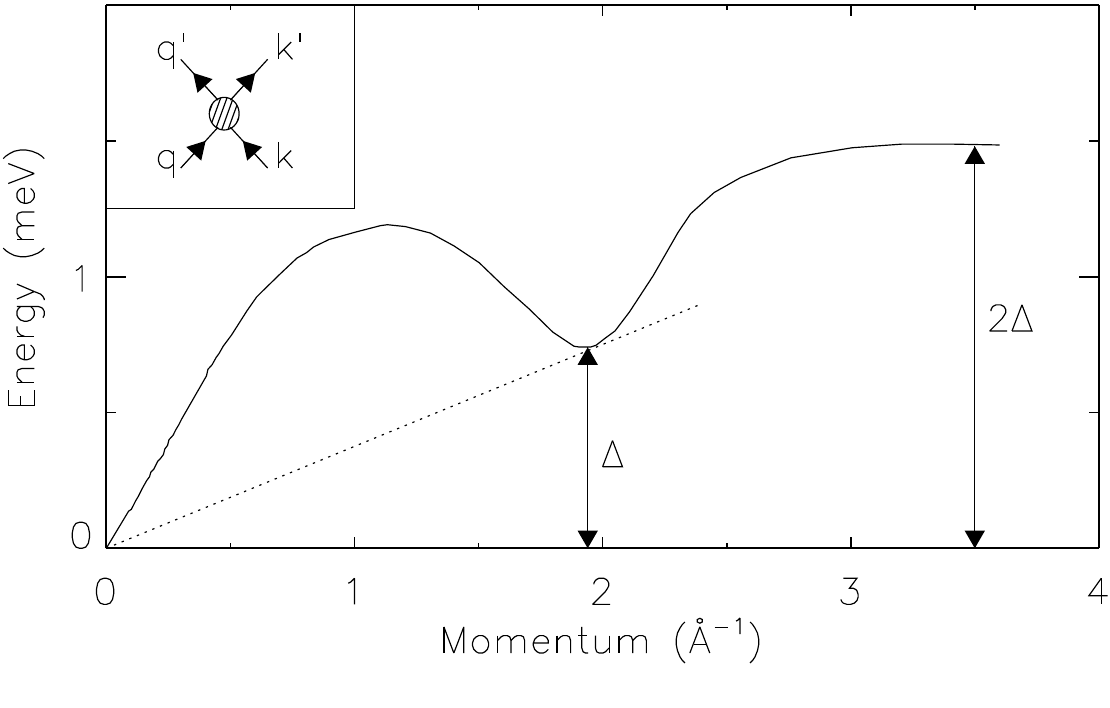}
	\caption{The phonon-roton dispersion curve of elementary excitations in superfluid helium. The roton minimum ($\Delta$) is located at q= 1.94 $\AA^{-1}$ and the curve terminates at twice the roton energy and momentum. The slope of the dotted line tangent to the curve corresponds to the minimum flow velocity below which the liquid behaves as a superfluid \cite{landau}. When the roton gap closes, the fluid ceases to be a superfluid. The inset sketches the Landau-Khalatnikov damping mechanism \cite{lk}: excitations with momentum $q$ acquire a finite lifetime through collisions with other excitations $k$. Independent of the details of this quasi-particle interaction, the resulting lifetime will be strongly temperature dependent as the weight of each diagram is proportional to the Bose population factor $1/(e^{\beta E_k}-1)$, with $\beta$ the inverse temperature $1/k_BT$ and $E_k$ the energy of the quasi-particle.}
	\label{disphel}
\end{figure}

Here, we show that the cause behind the opening up of a gap in the excitation spectrum is no mystery, but that it can be attributed fully to the temperature-dependent number of excitations falling below a critical value upon cooling. When this happens, the excitations change from being over-damped to propagating, resulting in the appearance of an energy gap as well as in a greatly enhanced thermal conductivity. This mechanism is independent of the exact nature of the excitations, instead it is entirely driven by the thermal population of the excitations via the well-known Landau-Khalatnikov \cite{lk} damping mechanism (see inset Fig. \ref{disphel}). We argue the validity of this mechanism in  URu$_2$Si$_2$  by making a direct comparison to the superfluid transition in helium-4.\\

The order parameter of the normal fluid to superfluid transition in helium-4 is well-known and the critical exponents associated with the transition have been verified \cite{shuttle} to be those of the 3D-XY model. Once the fluid is cooled down below T$_{\lambda}$= 2.172 K, the liquid stops boiling and alerts us to its new state by spontaneously emptying out of a beaker and leaking out of sealed containers through the tiniest of openings. The latter is a manifestation of the fluid being able to flow without friction through thin capillaries \cite{kapitza} whereas the lack of boiling bubbles is related to its exceptional heat conductivity.\\

The fraction of the liquid that can flow without friction is the superfluid fraction $\rho_s$; this fraction equals zero above  T$_{\lambda}$ and reaches 100 \% at T= 0 K \cite{glydebook}. This superfluid fraction is the order parameter of the superfluid phase. The fraction that gets dragged along in torsional experiments is referred to as the normal fluid fraction $\rho_n$, with $\rho_s$+$\rho_n$= 1. Experiments have revealed that the latter fraction is determined by the number of elementary excitations present at a given temperature. The dispersion of these elementary excitations, shown in Fig. \ref{disphel}, is commonly referred to as the phonon-roton dispersion curve. While $\rho_s$ reaches 100\% at T= 0 K, reflecting that there are no excitations present at absolute zero, the Bose-Einstein condensate fraction only reaches $\sim$7\% \cite{glyde}. This in contrast to an ideal, non-interacting Bose gas where the condensate fraction reaches 100\%, but which is not a superfluid as the ideal Bose gas quadratic dispersion does not have a minimum, non-zero slope. These well-known facts merely serve to illustrate that the condensate is a property of the ground state, but superfluidity is a property of the excited states. In particular, superfluidity (like superconductivity) materializes because there is a finite energy gap between the ground state and excited states.\\

Neutron scattering experiments have shown \cite{svensson} that the superfluid to normal fluid transition is marked by a very rapid decrease in lifetime of the elementary excitations, with these excitations continuously transforming into the density fluctuations that characterize normal liquids. \cite{svensson,wouterjltp,wouterbook} In other words, the phonon-roton excitations are (almost) the same density fluctuations as present in normal fluids, but with a much, much longer decay time. \cite{wouter00} The increase in decay time (decrease in damping) is entirely due to the disappearance of excitations upon lowering the temperature as the damping rate of excitations is determined by collisions rate between these quasi-particles. The very rapid changes observed in the damping rate of the elementary excitations just below T$_{\lambda}$ are the result of the roton-gap in the excitation spectrum opening up. \cite{wouter00,wouterbook}\\

In this paper we pursue the similarities between the hidden order transition and the $\lambda$-transition. In particular, we focus on the change in character of the excitation branches in going from the low-temperature to the higher-temperature phase and establish a one-to-one correspondence between the two systems based on the softening of the minima of the excitation curves in both systems. Doing so, we are able to provide a natural explanation for the most salient features associated with the hidden order transition.\\
\begin{table}[t]
 {
  \caption{Parameters characterizing the hidden order and the superfluid transitions.}
 \label{parameters}
 }
 {
  \begin{tabular}{c c c c c c}
    \hline
   &   \\[-2pt]
System & $T_{ord}$ & S($T_{ord}$) & Q$_{min}$ & gap(0 K) &$\Gamma$(18 K) \cite{niklowitz} \\[3pt]
  &K &Rln2 & nm$^{-1}$ &meV & meV\\[7pt]
      \hline\\[3pt]
$^4$He &  2.172  &1.08 &19.2  & 0.74 & 0.55 (SVP) \\[3pt]
 URu$_2$Si$_2$ ($Q_0$) &17.5 &0.22 &15 &2.2 & 2.3\\[3pt]
 URu$_2$Si$_2$ ($Q_1$)&17.5 &0.22 &9.1 &4.4 & 3.8\\[3pt]
 \hline\\ 
\end{tabular}
  }
\end{table}

This paper is organized as follows. In the next section we make a direct comparison between the entropy in the low temperature phases of $^4$He and  URu$_2$Si$_2$, showing that they behave virtually identically in warming up to the transition. We also compare the damping rates of the characteristic low temperature excitations (the roton excitation in $^4$He and the incommensurate $Q_1$ excitation in URu$_2$Si$_2$), demonstrating that both systems display a very similar increase in damping upon approaching the transition. In section III on the $\lambda$-transition in helium we review that the superfluid transition is marked by the point where the transitions between the two levels in $^4$He (the ground state and the first-excited state) become so frequent that the excitations go from being very well defined to a very heavily damped or even over-damped. This transition from well-defined (elementary) excitations to (almost) diffusive transport is complete at the transition point.  In section IV we show that identical considerations hold for the hidden order transition in URu$_2$Si$_2$, and that most of the puzzling features associated with hidden order can be directly associated with a change from well-defined propagating excitations in the hidden-order phase to diffusive, over-damped excitations in the Kondo lattice phase above $T_{HO}$. In our discussion section we argue provided that the lowest excitations in URu$_2$Si$_2$ are between two  singlet states, that the sharp increases in resistivity and thermal conductivity on cooling through the hidden order transition are a direct consequence of the changes in the damping mechanism of the excitations. From our comparison with helium we argue that these changes in damping mechanism are entirely attributable to the number of quasi-particles excited at a given temperature, with their damping rates (lifetimes) governed by the number of collisions between the quasi-particles.

\section{Global comparison between $^4$He and URu$_2$Si$_2$}
\begin{figure}[t]
\begin{center} 
\includegraphics*[viewport=160 130 480 400,width=85mm,clip]{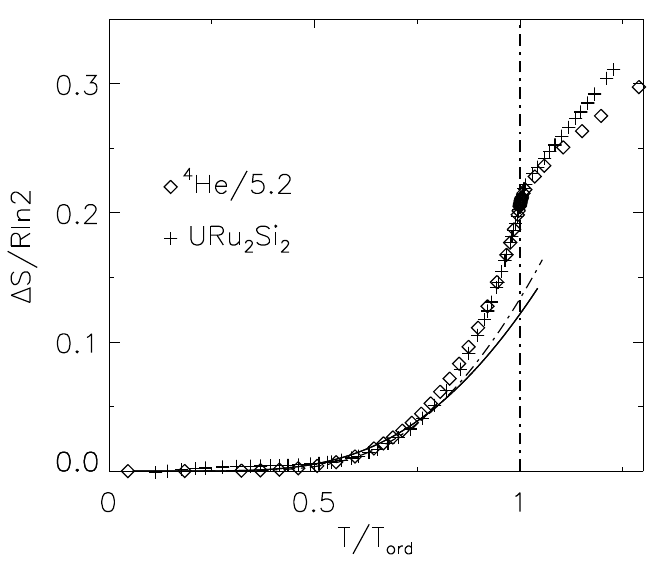}
\end{center}
	\caption{The measured entropy for URu$_2$Si$_2$\cite{dijk97} (with the value at T= 2 K subtracted to eliminate the superconducting transition) and $^4$He \cite{donnely}. A vertical scale factor of 5.2 has been applied to the helium data, and the results have been plotted versus reduced temperature. The dashed-dotted line is the predicted entropy for $^4$He using the T= 0 K dispersion for the calculation, the solid curve (largely coinciding with the dashed curve)  is the same for  URu$_2$Si$_2$ using the dispersion from reference \cite{broholm}.}
	\label{entropy}
\end{figure}

Both the hidden order transition in URu$_2$Si$_2$ at 17.5 K and the superfluid transition in $^4$He at 2.172 K are marked by a $\lambda$-shaped anomaly in the specific heat c \cite{dijk97,glydebook}. Integrating the c/T curves from zero to T yields the entropy of the systems, which we display in Fig. \ref{entropy}. In order to make a direct comparison between the two systems, we apply scaling both in this figure and in subsequent figures. We give the relevant parameters used in our scaling procedures in Table \ref{parameters}. In addition, in order to bring the similarity between the two systems to the fore, we have removed the entropy change that is due to the superconducting transition at 1.5 K \cite{palstra} from the URu$_2$Si$_2$ data: $\Delta S= S - S(T= 2 K)$. This is justified as the superconducting transition occurs at a temperature well below any changes associated with the hidden-order transition. It is clear from Fig. \ref{entropy} that the temperature evolution of the entropy upon approaching the transition from below is very similar in shape. However, the entropy of helium at the transition is roughly Rln2 \cite{donnely}, that of  URu$_2$Si$_2$ is lower by a factor of five \cite{dijk97}, indicating that the hidden order transition does not correspond to a complete freezing out of a degree of freedom; most of the entropy has already been shed at higher temperatures.\\

\begin{figure}[t]
\begin{center}
\includegraphics*[viewport=190 135 620 390,width=85mm,clip]{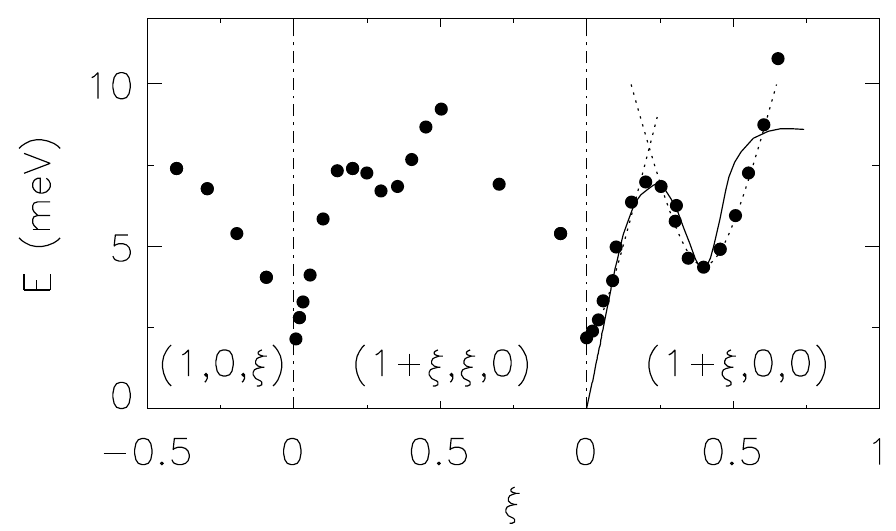}
\end{center}
	\caption{The excitation energies (symbols) along high-symmetry directions indicated in the plot for URu$_2$Si$_2$ as measured by Broholm {\it et al.} \cite{broholm} in the hidden-order phase. The solid line is the phonon-roton dispersion curve plotted for comparison and scaled to make the roton coincide with the $Q_1$-minimum of the  URu$_2$Si$_2$ dispersion. The dotted lines are the approximations by Williams {\it et al.} \cite{williams} for the two minima $Q_0$= (1,0,0) and $Q_1$= (2-0.6,0,0).}
	\label{disp}
\end{figure}

A similar close agreement between the two systems is found when looking at the damping rates of the excitations at the minimum of the dispersion curves. We reproduce the dispersion of  URu$_2$Si$_2$ in the hidden order phase as measured by Broholm {\it et al.} \cite{broholm} in Fig. \ref{disp} for three crystallographic directions. The two lowest minima of the dispersion occur at q=(1,0,0) and at (0.6,0,0) with gaps of 2.2 meV and 4.5 meV, respectively. We refer to their positions in reciprocal space with $Q_0$ and $Q_1$. We also show the direction-independent curve of elementary excitations \cite{wouter00} (Fig. \ref{disphel}) of the superfluid as a solid line, scaled so that the minimum of the $^4$He dispersion coincides with the minimum at $Q_1$.\\

\begin{figure}[t]
	\includegraphics*[viewport=120 100 570 430,width=85mm,clip]{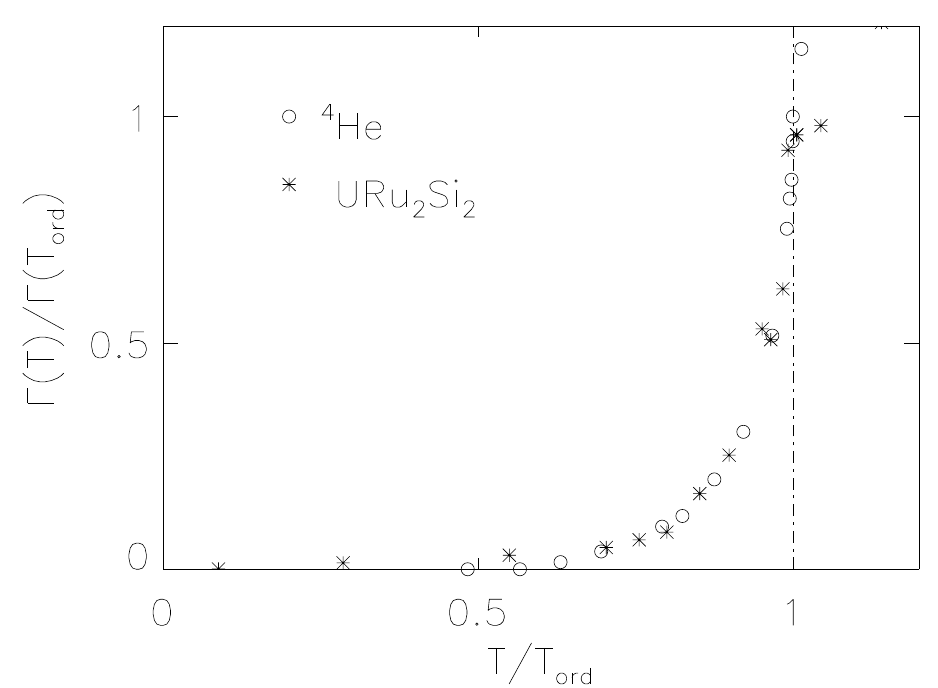}
	\caption{The damping rates at the roton minimum \cite{wouterjltp} and the excitations at $Q_1$ \cite{bourdarot10} versus reduced temperature, made to coincide at $T_{ord}$. The temperature evolution of both rates is essentially identical.}
	\label{damping}
\end{figure}

Neutron scattering experiments \cite{bourdarot10,niklowitz,svensson} have revealed that the sharp (in energy) excitations in $^4$He and  URu$_2$Si$_2$ broaden considerably upon approaching the transition from below, with the largest increases occurring close to the transition. This is shown in Fig. \ref{damping}. At the lowest temperatures, the damping rates are very small, implying long life times for the excitations. Long-lived excitations are referred to as elementary excitations, or well-defined quasi-particles. It is clear from this figure that both systems share a similar evolution in the decay rate (inverse life time) of their excitations, and that very rapid changes occur in the damping mechanism close to the transition.\\ 

In the next section we review the superfluid state with the aim of facilitating the comparison between the two systems. Unlike URu$_2$Si$_2$, superfluid helium is a relatively simple system with the only signal that shows up in neutron scattering experiments being the density fluctuations that are created (or absorbed)  by the neutron. In addition, the order parameter of the $\lambda$-transition is well known \cite{shuttle} and characterized as being the superfluid fraction. Thus, the aim of the next section is catalogue exactly what is changing between the normal fluid and the superfluid phase in helium so that we can then utilize the similarity between $^4$He and URu$_2$Si$_2$ to learn more about the hidden order transition.

\section{The ${\lambda}$-transition in $^4$He}
We first elucidate how the $\lambda$-transition in $^4$He is linked to the opening up of a gap in the excitation spectrum as there exists the common misconception that the spectacular properties of superfluid helium are the result of the presence of the Bose-Einstein condensate. Most simply put, helium can flow without friction because a minimum flow velocity is required in order to transfer the energy and momentum of the liquid flow into internal degrees of freedom of the liquid, that is, creating elementary excitations. This minimum requirement is present because of a finite energy gap between the ground state and the excited states. This gap in turn is present because of the Bose statistical nature of the $^4$He atoms, not because of the presence of a Bose-Einstein condensate. We refer to the textbook by Feynman \cite{feynmanbook} for the basic arguments behind these facts.\\

We show the details of the elementary excitation spectrum of superfluid helium in Fig. \ref{disphel}. This phonon-roton excitation curve captures the sound-like excitations at small momentum transfers $\hbar q$, the roton minimum with energy gap $\Delta$ at wavelengths $\lambda=2\pi/q=d$ corresponding to the average inter-atomic separation $d$, followed by a flat part of the dispersion that terminates at twice the roton momentum at a level of 2$\Delta$. The minimum flow velocity is given by the tangent to the dispersion curve (dashed curve in Fig. \ref{disphel}) which reaches its smallest value close to the roton minimum. Experiments where the liquid was made to flow through very small orifices \cite{varoquaux} (in order to impede the formation of topological excitations) have shown that this minimum slope does indeed correspond to the critical velocity as predicted \cite{landau} by Landau. Thus, it is the presence of the roton gap that ensures the property of superfluidity, making it clear that superfluidity is a property of the excited states.\\

The order parameter of the superfluid transition is the superfluid fraction $\rho_s$, being the fraction of the liquid that can flow through thin capilaries without friction. The normal fluid fraction $\rho_n$ (with $\rho_n$+$\rho_s$=1), being the part of the liquid that is dragged along in torsional balance experiments, is a measure of how many elementary excitations are present. Therefore, the order parameter is not linked to the condensate fraction of the 7\% of the atoms \cite{glyde} that have formed a Bose-Einstein condensate in the ground state, instead it is determined by the shape of the elementary dispersion curve. While not needed in this paper, we note that the universality class of the superfluid transition has been confirmed by specific heat experiments \cite{shuttle} on board the space shuttle in micro-gravity conditions in order to determine the critical exponent $\alpha$. Thus, the order parameter of superfluid helium holds no mystery, unlike the hidden order transition in  URu$_2$Si$_2$.\\

On a microscopic scale, the superfluid to normal fluid transition is marked by a complete filling in of the roton gap. Neutron scattering experiments \cite{svensson,wouterjltp} very close to T$_{\lambda}$ have revealed that the transition from the elementary excitations of the superfluid (Fig. \ref{disphel}) to the normal fluid density fluctuations is a continuous one, with most of the changes occurring within 0.05 K of T$_{\lambda}$. The excitations are seen to become heavily damped on raising the temperature as a direct result of Landau-Khalatnikov damping (inset Fig. \ref{disphel}), as shown in Fig. \ref{damping} for the roton excitation. For the roton, the damping rate becomes comparable to the energy gap, with the result that the gap disappears, and the fluid ceases to be a superfluid as a direct consequence. Note that this increase in damping resulting in a change in character cannot be avoided: it is something that necessarily must happen at a given temperature given the increase in the number of quasi-particles that are thermally excited. In helium, the elementary excitations of the superfluid phase change completely in character because of the increase in damping and transform into the ordinary density fluctuations of normal fluids. This statement was verified \cite{wouter00} through second-order perturbation theory by using ordinary density fluctuations as basis functions and linking them to multi-particle states. This perturbation expansion was successful in reproducing \cite{wouterbook,wouter00} the entire phonon-roton dispersion curve, including the intensity of the excitations in neutron scattering experiments as well as the termination point of the dispersion at twice the roton momentum. Thus, on a microscopic scale, the superfluid transition can be fully understood as being the consequence of the unavoidable filling in of the roton gap.\\

In the next section we argue that the hidden order transition in URu$_2$Si$_2$ is of a similar nature in the sense that the same excitations are present both above and below the transition, but that a marked change in their damping rates and character (reflecting the rapid increase in the number of excitations present) accounts for the difference between the hidden order phase and the Kondo-shielded metallic phase above the transition.

\section{The hidden order transition in URu$_2$Si$_2$ }
In this section we argue that the hidden order transition in URu$_2$Si$_2$ is associated with the two lowest levels of the electronic subsystem, with the levels becoming degenerate above the hidden order transition because of thermal population effects. We demonstrate that this change in character accounts for the most salient features of the hidden-order transition, and provided the two levels are magnetic singlets, it also accounts for the differences in thermal conductivity and electrical resistivity between the hidden order state and the Kondo lattice state above $T_{HO}$.

\subsection{The hidden order phase of URu$_2$Si$_2$ }

URu$_2$Si$_2$ above $T_{HO}$ is a metal that exhibits Kondo screening \cite{mydosh11} as evidenced by the observed coherence in the resistivity and the enhanced electron effective mass just above  $T_{HO}$ resulting in a strongly enhanced coefficient of the linear term of the specific heat ($\gamma$= 160 mJ/mol.K$^2$ \cite{mydosh11}). Cooling down through $T_{HO}$ sees a jump in the specific heat curve ($\Delta$c$\approx$ 300 mJ/K) over a narrow temperature interval ($\Delta$T $\approx$ 0.2 K), with an overall $\lambda$ shape for the entire curve. At the same time and quite surprisingly, a significant increase is observed, upon entering the HO-phase, in the thermal conductivity as well as in the resistivity.\\

Neutron scattering experiments have shown that a weak anti-ferromagnetic (AF) Bragg peak appears \cite{broholm87} at $T_{HO}$, corresponding to a tiny fraction of the U-moment ($\sim$ 0.02 $\mu_B$). This Bragg peak is not a true Bragg peak in the sense that the ordering is not truly long range, but is limited \cite{broholm} to about 50 nm. The size of this ordered moment is so small that AF ordering cannot correspond to the observed changes in specific heat at $T_{HO}$ as a moment that small must already have lost most of its entropy during the Kondo-shielding process. \cite{sikkema} In addition, the inferred size of the ordered moment is strongly sample dependent \cite{mydosh20} and non-uniform \cite{matsuda} throughout a sample, suggesting that this small moment phase might not be intrinsic. Inelastic neutron scattering experiments showed that magnetic excitations are present in the HO-phase \cite{broholm87,broholm}, and that these excitations are gapped throughout the entire Brillouin zone, with excitation minima observed at the AF-position $Q_0$ ($\Delta$= 2.2 meV) and at incommensurate positions $Q_1$ ($\Delta$= 4.4 meV). We reproduce the earliest measurements of the dispersion curve in Fig. \ref{disp}.\\

Based on the details of the scattering process and on the (lack of) field dependence of the excitation energies, these excitations were identified \cite{broholm} as transitions between two magnetic singlet states. Thus, these excitations are propagating fluctuations of the electronic subsystem. We note that transitions between singlet states that do not have a net magnetic moment will still be observed in magnetic neutron scattering experiments and in susceptibility measurements: Whatever the exact nature of the two singlet states (such as the states considered by Sikkema {\it et al.} \cite{sikkema}), the operator J$^z$ will induce transitions between them. In magnetic neutron scattering experiments this corresponds to non spin-flip scattering, or longitudinal excitations. As far as we are aware, this is indeed what has been observed for the magnetic inelastic cross-section. In susceptibility measurements, the transitions show up as van Vleck terms \cite{fazekas}. Thus, even though no net magnetic moment needs to be present in URu$_2$Si$_2$, it will still have a magnetic response. We note that recently the singlet character of the lowest lying states has been confirmed \cite{severing} through inelastic Xray scattering experiments.\\ 

These spatially extended singlet states, representing propagating excitations, acquire their dispersion through the RKKY-interaction J$_{\vec{q}}$. \cite{broholm} The RKKY-interaction strength is strongest at $Q_0$ and $Q_1$ as evidenced by the dispersion minima at these wave vectors. Scattering experiments \cite{wiebe07} have shown that this interaction still leads to a strong scattering rate at $Q_1$ above $T_{HO}$, but now without an energy gap being present.\\

The opening up of an excitation gap between the two states at $T_{HO}$ has been observed in many types of experiments, perhaps most convincingly \cite{aynajian} so by spectroscopic imaging scanning tunneling microscopy (SI-STM). These experiments confirm Kondo screening above $T_{HO}$ where in the absence of an energy gap the two singlet states are degenerate, allowing for the Kondo shielding mechanism to materialize. They also show that below $T_{HO}$, the degenerate band is split into two separate bands, and the Kondo shielding process is thwarted.\\

Finally, experiments have been performed as a function of magnetic field \cite{dijk97} and hydrostatic pressure. \cite{villaume} The field dependent measurements revealed that $T_{HO}$ is lowered somewhat by moderate fields, and the hidden order phase can be destroyed \cite{levallois} by applying large fields (in excess of 30 T). Application of hydrostatic pressure \cite{villaume} results in a modest increase in $T_{HO}$ up to a critical pressure. At the same time, the ordered AF moment increases with applied pressure, until a transition from the HO-phase to a true long-range ordered AF-state is reached. Below T$\approx$ 1.5 K superconductivity appears \cite{palstra} in the HO-phase, however, we will not address this in this paper.

\subsection{A more detailed comparison of the hidden order and the $\lambda$-transitions}

In this subsection we perform a slightly more detailed comparison between the two systems in order to show that both systems are two level systems, with the excited level becoming increasingly more populated upon raising the temperature, leading to a closure of the gap between the two levels at the transition.\\

In Fig. \ref{entropy} we already did a direct comparison between the entropy of both systems, we now show that this entropy can be predicted numerically based on the measured dispersion curves. In Fig. \ref{entropy} we show the expected entropy for $^4$He based on the dispersion curve as the dashed-dotted curve, using standard statistical methods \cite{ashcroft} to relate the dispersion to the specific heat. Of course, it was the measured specific heat of helium that led Landau \cite{landau} to make his now famous prediction of the phonon-roton curve, well before neutron scattering experiments revealed its correctness. The measured entropy and predicted value start to deviate at elevated temperatures (T$>$ 1.3 K) when the excitation energies start to become appreciably smaller because of the increased damping. Including the measured softening \cite{bedell} of the roton excitation would extend the agreement between the curve and the experimental entropy to slightly higher temperatures, but (of course) the region right at the transition cannot be adequately captured without including power-law behavior.\\

We perform an identical procedure \cite{ashcroft} for URu$_2$Si$_2$  and show the results in Fig. \ref{entropy} (solid curve). We use the dispersion curve from Fig. \ref{disp} in combination with the results from other experiments such as those by Wiebe {\it et al.} \cite{wiebe04} where it was shown that most of the scattered intensity appears in a ring in the ab-plane centered around $Q_0$ with a radius of 0.4 reciprocal lattice units. In Fig. \ref{disp} there are two points of the dispersion visible along this ring: $Q_1$ and $q$=(1.3,0.3,0). Note that the gap energy is higher at the latter point, although still at a local minimum. We approximate the entire dispersion curve by parabolic minima located on this ring as well as a by an additional single minimum located at $Q_0$. We use the measured experimental parameters to extend these points throughout the entire Brillouin zone. Examples are given by the dotted parabolic curves in Fig. \ref{disp} where we have used the relevant velocities determined by Williams {\it et al.} \cite{williams} (v= 24 meV{\AA} in plane and v= 32 meV{\AA} parallel to the c-axis). The results of our calculation are shown by the solid curve in Fig. \ref{entropy}. Note that we have not allowed for the gap to soften upon approaching the transition in these calculations. Van Dijk {\it et al.} \cite{dijk97} have already shown that by allowing for a softening of an effective gap (with a complete softening at $T_{HO}$) that the entropy curve can be reproduced over a much larger temperature range. Overall, the data in this figure demonstrate that the entropy changes as a function of temperature in  URu$_2$Si$_2$ are to be ascribed to the changes in population of the excited states, similar to the case for helium.\\

Note that in our calculations for both systems, the entire dispersion curve has been used as input. In the case of helium, the phonon contributions to the specific heat yield a $\sim$T$^3$ term which is overtaken in importance by contributions from excitations near the roton minimum at around 0.7 K. At the superfluid transition, the phonon contributions are significantly less important that the rapidly increasing number of softening roton excitations. A somewhat similar situation takes place in URu$_2$Si$_2$. Even though the $Q_0$ gap is lower in energy than the $Q_1$ gap, there are four $Q_1$ points in the first Brillouin zone with the result that the $Q_1$ contributions account for roughly double the $Q_0$ contribution at the hidden-order transition.\\

Another striking similarity between the two systems is the sharp increase in damping rate (decrease in lifetime) upon approaching the transition from below. We already showed this correspondence in Fig. \ref{damping}. This rapid increase in damping is observed both at $Q_0$ and at $Q_1$. Therefore, in a manner identical to superfluid helium, approaching the transition from below is marked by a very rapid increase in damping rate, resulting in a disappearance of the gap between the ground state and the first excited state. In helium, this increase in damping is driven solely by an increase in the number of quasi-particles present resulting in an increased collision rate between them (see Fig. \ref{disphel}), not by an underlying change in the {\it undamped} gap value. By extension, it stands to reason that the hidden order transition is caused by the same rapid increase in the number of quasi-particles present, with the damping mechanism given by the number of collisions between these quasi-particles as depicted in the renormalized four-vertex diagram shown in Fig. \ref{disphel}.\\
 
\begin{figure}[t]
\begin{center}
\includegraphics*[viewport=110 60 530 515,width=85mm,clip]{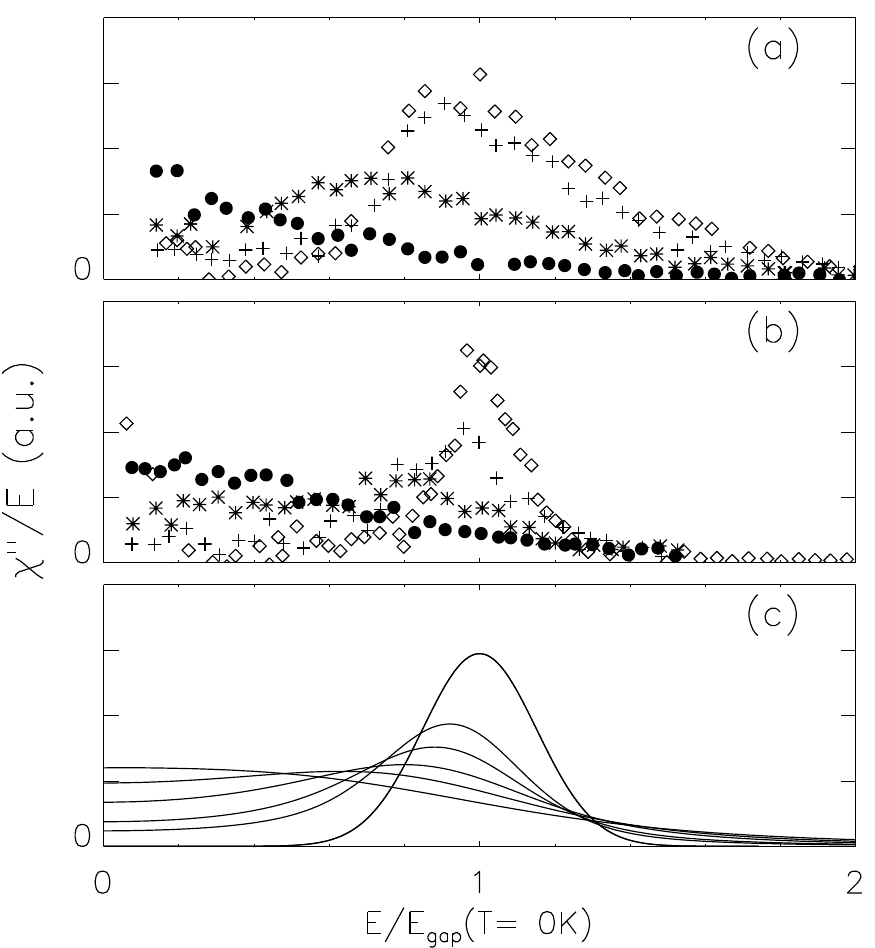}
\end{center}
	\caption{Neutron scattering data on  URu$_2$Si$_2$ for $Q_0$ (a) and $Q_1$ (b) plotted as $\chi"$/E (to eliminate the Bose population factor) and reproduced from Niklowitz et al. \cite{niklowitz} (diamonds: T= 3 K; pluses: 12.5 K; asterisks: 16 K; filled symbols: 18 K). Without any data analysis besides the subtraction of a temperature-independent flat background, the data already clearly show the transition from well-defined excitations to heavily-damped quasi-elastic features upon leaving the hidden-order phase, in a manner strongly reminiscent of $^4$He (c). The $^4$He-data have been reproduced from \cite{wouterjltp} and convoluted with a Gaussian in order to mimic the energy resolution of the URu$_2$Si$_2$ experiments. The data are shown for temperatures of 1.1 K (highest peak), 1.9 K, 2.0 K, 2.1 K, 2.15 K, and 2.2 K (quasi-elastic feature), demonstrating the smooth transition from propagating (gapped) to non-propagating as the temperature is raised through the $\lambda$-transition at 2.172 K. The lines shown in this panel are the (essentially perfect) damped-harmonic oscillator fits detailed in \cite{wouterjltp}.}
	\label{softmode}
\end{figure}

In order to emphasize this point of view, in panel (c) of Fig. \ref{softmode} we reproduce the temperature dependence of the response of helium at the roton minimum when it goes through the superfluid transition as measured by neutron scattering experiments. \cite{wouter00} The signal seen in neutron scattering experiments is distorted because of the presence of the Bose population factor; we therefore remove this factor and plot the Fourier transform of the relaxation function \cite{relaxation} instead ($\chi(E)"/E$ with $\chi"$ the imaginary part of the dynamic susceptibility) in order to view the temperature dependence of the poles of the dynamic susceptibility. It is clear from this figure that rapid changes occur in the response, with most changes taking place close to the transition temperature. We see that the (resolution-broadened) well-defined low temperature excitations become increasingly more damped, until they become, for all practical purposes, quasi-elastic non-propagating features at the transition. In helium this marks the transition from superfluid to normal fluid behavior, and it is driven entirely by the increase in damping rate (Fig. \ref{damping}), which in turn  is driven entirely by the number of excitations present. Both above and below T$_{\lambda}$ neutron scattering measures density fluctuations, but these fluctuations become less and less damped upon lowering the temperature, until at the very lowest temperature the damping is so reduced that these density fluctuations acquire the character of elementary excitations.\\

The response of URu$_2$Si$_2$ upon going through the hidden-order transition (Fig. \ref{damping} and panels (a) and (b) of Fig. \ref{softmode}) is very similar to that of helium. We see a marked increase in damping rate upon raising the temperature (Fig. \ref{damping}), with the most rapid increase occurring close to the transition. Above the transition, the excitations have become over-damped (visible to the naked eye in Fig. \ref{softmode}), implying that there are now two energy levels in the system with the same energy. When the temperature is lowered through $T_{HO}$ then the excitations become under-damped, and they can now propagate through the system. Thus, rather than having a diffusive character (over-damped), they have now become propagating waves. In the discussion section we detail how this change from over-damped to under-damped affects Kondo screening and the Fermi surface, and how the change from diffusive behavior to propagating is reflected in the thermal conductivity of  URu$_2$Si$_2$.\\  

\section{discussion}
In the previous section we have established the close connection between the normal-fluid to superfluid transition in $^4$He and the hidden order transition in URu$_2$Si$_2$. We have implied that because of this similarity, the hidden order transition is one that is driven by the thermal population of the excitations to the point where the collision rate between the excitations becomes so large upon heating that the gap between the two singlet states disappears, and the system becomes a Kondo-shielded metal. However, we have to look into the many other aspects that have been discovered about the HO-phase to ascertain that this scenario stands up to scrutiny. We show in this section that from a qualitative point of view this is indeed the case and end with some suggestions for future experiments as well as some speculative remarks.

\subsection{Transport}
Upon lowering the temperature through the hidden order transition, there is a noticeable increase \cite{behnia,sharma} in resistivity as well as in thermal conductivity. A large fraction of the carriers is lost, and  URu$_2$Si$_2$ becomes an almost compensated metal below $T_{HO}$ with about 0.1 holes per uranium ion. \cite{mydosh11}  As such, the increase in thermal conductivity is even more astounding. However, the scenario sketched in the preceding sections offers a natural interpretation for this.\\

The increase in thermal conductivity can be interpreted as a transition from the excitations being over-damped (diffusive) to becoming propagating. Excitations carry energy, and as such, a transition from diffusive to propagating would herald a large increase in the speed at which energy can be transported. Since the number of excitations decreases precipitously with decreasing temperature, this increase in conductivity would be restricted to temperatures just below $T_{HO}$ but initially, the increase will easily overcome the loss of conduction electrons when the Fermi-surface shrinks (see next paragraph). Note that this increase in thermal conductivity is independent of the exact nature of the excitations in the HO-phase, whether they are associated with magnetic singlets or not.\\

Above $T_{HO}$, Kondo shielding takes place and the otherwise localized U-electrons partake in the transport mechanism. In  URu$_2$Si$_2$ above $T_{HO}$, the Kondo screening mechanism is not facilitated by a magnetic ground state doublet, but rather by the two degenerate magnetic singlet levels. Once this degeneracy is lifted, then Kondo shielding is impeded because of the opening up of the gap between the singlets. In our scenario, this happens when the number of excitations drops enough for the gap to open up in the dynamic susceptibility. As such, the increase in resistivity below $T_{HO}$ simply reflects the disappearance of the Kondo shielding mechanism. While this is not a new insight as an arrested Kondo effect has already been associated with the hidden order transition \cite{haule}, where we differ with the literature is in the reason for this gap opening up as being caused by a thermally driven decrease in the number of quasi-particles and consequently by a reduced collision rate between them, rather than by any fundamental change in the nature of the system.\\
 
\subsection{Order parameter}
The order parameter for superfluid helium is a complex function whose amplitude yields the temperature dependence of the superfluid fraction and whose phase captures the extent of the long range correlations that allow helium to creep out of a beaker. Thus, the magnitude of the order parameter captures an aspect of the excited states. Below T$_{\lambda}$, the ground state of helium also exhibits marked changes with the appearance of the Bose-Einstein condensate that sees a small fraction \cite{glyde} ($\sim$ 7\%) of the He-atoms condense into the zero momentum state. However, the condensate fraction does not serve as the order parameter. Given the similarity between helium and URu$_2$Si$_2$, we discuss the nature of the 'hidden order' parameter and the appearance of the small moment elastic Bragg peak that materializes below $T_{HO}$.\\ 

\begin{figure}[t]
\begin{center}
\includegraphics*[viewport=105 145 580 430,width=85mm,clip]{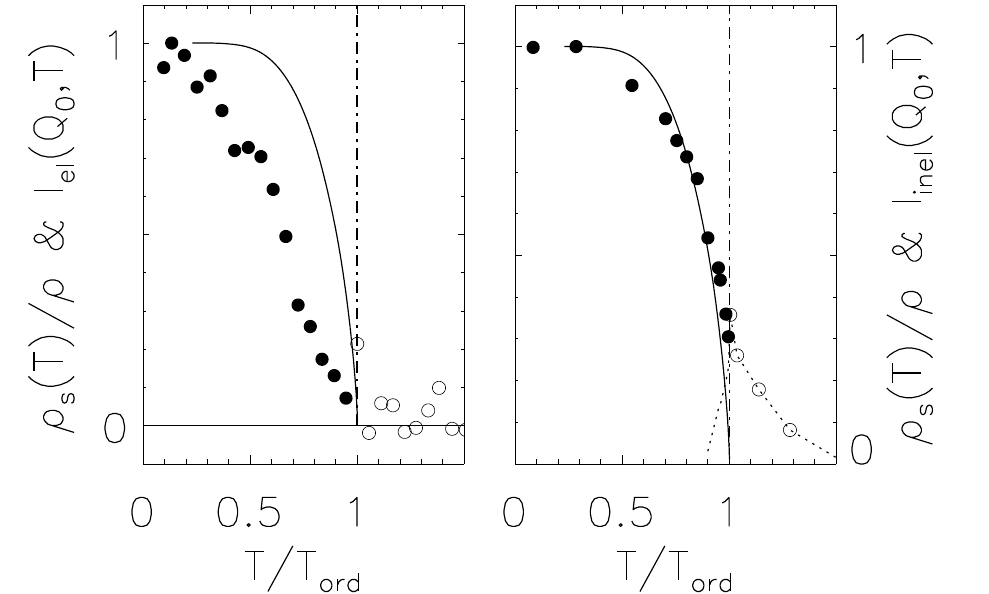}
\end{center}
	\caption{Comparison of the elastic (left panel) and inelastic intensities (right panel) in  URu$_2$Si$_2$ (filled circles: T $<$ $T_{HO}$, open circles:  T $>$ $T_{HO}$) with the superfluid density in $^4$He (solid line). The elastic intensity displays the temperature evolution of the (1,0,0) incipient Bragg peak (as the order is not truly long range \cite{broholm}), the inelastic intensity is the integration of the imaginary part of the dynamic susceptibility for 0 $<$ E $<$ 6.3 meV (reproduced from Bourdarot {\it et al.} \cite{bourdarot10}) The left panel shows that the Bragg peak does not coincide with the temperature evolution of the $^4$He order parameter. The integrated inelastic intensity (right panel) follows the $^4$He order parameter much more closely, and can be made to coincide by assuming the presence of critical scattering (dotted curve).}
	\label{order}
\end{figure}

We plot the superfluid density and the intensity of the (1,0,0)-Bragg peak in URu$_2$Si$_2$ as a function of T/T$_{ord}$ in the left panel of Fig. \ref{order}. It is clear from this figure that the temperature dependence of the two quantities is fundamentally different. We observe a much closer agreement (right panel)  between $\rho_s(T)/\rho$ and the integrated intensity of the inelastic part of the (1,0,0)-excitations. This correspondence between the excited states in URu$_2$Si$_2$ and the superfluid density (a measure of how many excitations are no longer present) is in agreement with our proposed scenario that the transition is driven by changes to the excited states caused by an increased number of excitations present.\\

Even though the number of excitations present would determine the transition temperature, this number by itself is not the order parameter. In superfluid helium, it is the number of excitations present multiplied by their mass equivalent (as we are describing superflow of particles), with the mass equivalent also being temperature dependent. \cite{landau} It is unclear, assuming that the similarity with helium carries all the way, what property needs to be multiplied with the number of excitations in order to arrive at the order parameter in URu$_2$Si$_2$. While the excitations carry energy and momentum, they do not carry a magnetic moment. They probably also do not carry charge as they are (most-likely) propagating electron-hole excitations (excitons, or transitions between the top of the valence band and the bottom of the conduction band \cite{butch}). Therefore, the spin and charge attributes of the electron would likely not factor into the order parameter, in line with the observed lack of symmetry change of the excitations on going through the hidden order transition \cite{butch}.\\

We note that the comparison with superfluid helium can only go so far: in $^4$He, the density fluctuations are the only excitations left and as such, reducing their number by cooling down sees them evolve into elementary excitations. In contrast, URu$_2$Si$_2$ still has many degrees of freedom left and we expect the excitations in the hidden-order phase to remain fractionally damped, not allowing for the complete separation between the equivalent of the normal fluid and superfluid component as energy transfer between the two remains possible. Therefore, it is quite possible that the HO-order parameter does not reach its full extent on cooling down.\\ 

Overall, the connection between the amplitude of the order parameter and the number of excitations present in URu$_2$Si$_2$ appears to agree with all experimental observations. Applying a magnetic field lowers the hidden-order transition temperature; measuring the field-dependent entropy revealed \cite{dijk97} that the entropy at $T_{HO}$ is independent of field. This is what is expected when the transition is driven by the number of excitations present for a given gap energy. When the $Q_1$ gap energy increases, as it does with applying pressure, then we see that  $T_{HO}$ also increases. Again, this is as expected since now more excitations should be present before the damping rate matches the gap energy. In all, it appears that the incommensurate $Q_1$-gap energy is the determining parameter, similar to how the roton energy is the determining parameter in helium, with the thermally-driven opening up of the gap responsible for the $\lambda$-anomaly in the specific heat at $T_{HO}$.\\

\subsection{The origin of the entropy at $T_{HO}$}
We showed in Fig. \ref{entropy} that the superfluid transition is associated with $\sim$Rln2 in entropy, but the hidden order transition is only of the order of 0.2 Rln2. \cite{butch} This raises questions both as to its smallness as well as to its largeness. On the one hand, in our sketched scenario, all singlet levels are partaking so we would expect the associated entropy to be 2Rln2 (Rln2 for each of the two 5f2-electrons). On the other hand, the system above the hidden-order transition is a Kondo shielded system with a Kondo temperature of the order of 70 K, so we expect all the local moment degrees of freedom to have frozen out at $T_{HO}$ \cite{sikkema}, leaving no entropy to be shed in the hidden-order phase. That is, when the system goes from being Kondo shielded to having an energy gap greatly exceeding the sample temperature, then there is no net change in available degrees of freedom (merely a rearrangement) and we would expect zero entropy to be associated with the transition, removing the signature of the transition from the specific heat curve.\\

\begin{figure}[t]
\begin{center}
\includegraphics*[viewport=110 140 550 410,width=85mm,clip]{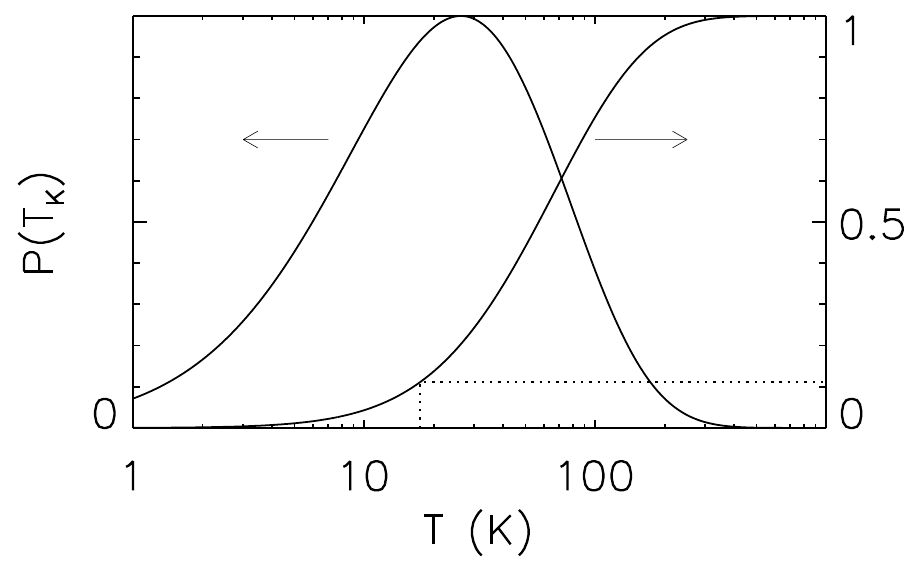}
\end{center}
\caption{Instantaneous distribution of Kondo shielding temperatures (left axis) and resulting fraction of ions where the formation of a magnetic singlet is favored (right axis) for URu$_2$Si$_2$. Note the logarithmic temperature axis. The average Kondo temperature was chosen to be 75 K, the conduction bandwidth was taken to be 0.4 eV based on theoretical estimates \cite{elgazzar}, and the Debye-Waller factor for the U-Ru distance was fixed at 0.006 nm based on experimental observations \cite{RuU}. The dotted lines indicate that at $T_{HO}$= 17.5 K 11\% of the moments are not Kondo shielded at any instance in time. Figure adapted from reference \cite{bretana}.} 
\label{kondo}
\end{figure}

A very similar issue was recently resolved \cite{bretana1,bretana} in Kondo lattice systems and accounts for the entropy magnitude of 0.2 Rln2. Whereas the average Kondo temperature is well in excess of $T_{HO}$, the zero-point motion of the ions induces a broad instantaneous distribution of shielding temperatures. For instance, a 1\% distribution in Ru-U separation results in a 50\% distribution of shielding temperatures owing the the exponential sensitivity of the Kondo scale to inter-ionic separations \cite{bretana}. As long as the electronic time scales are much faster than the ionic time scales, this instantaneous distribution is experienced by the electrons as a static distribution. We reproduce the Kondo shielding temperature distribution in Fig. \ref{kondo}. Note that there is only one adjustable parameter in this distribution, namely the average Kondo temperature. The other two parameters (conduction bandwidth and Debye-Waller factor) are known from experiments and calculations. It can be seen in this figure that at $T_{HO}$ roughly a 10\% fraction of the moments is not Kondo shielded at any given time. This would account for the $\sim$0.2 Rln2 entropy still available at $T_{HO}$ (there being two electrons per U ion), rendering our scenario for the transition self-consistent.\\

\subsection{Pressure dependence}

\begin{figure}[t]
	\begin{center}
		\includegraphics*[viewport=0 0 620 440,width=85mm,clip]{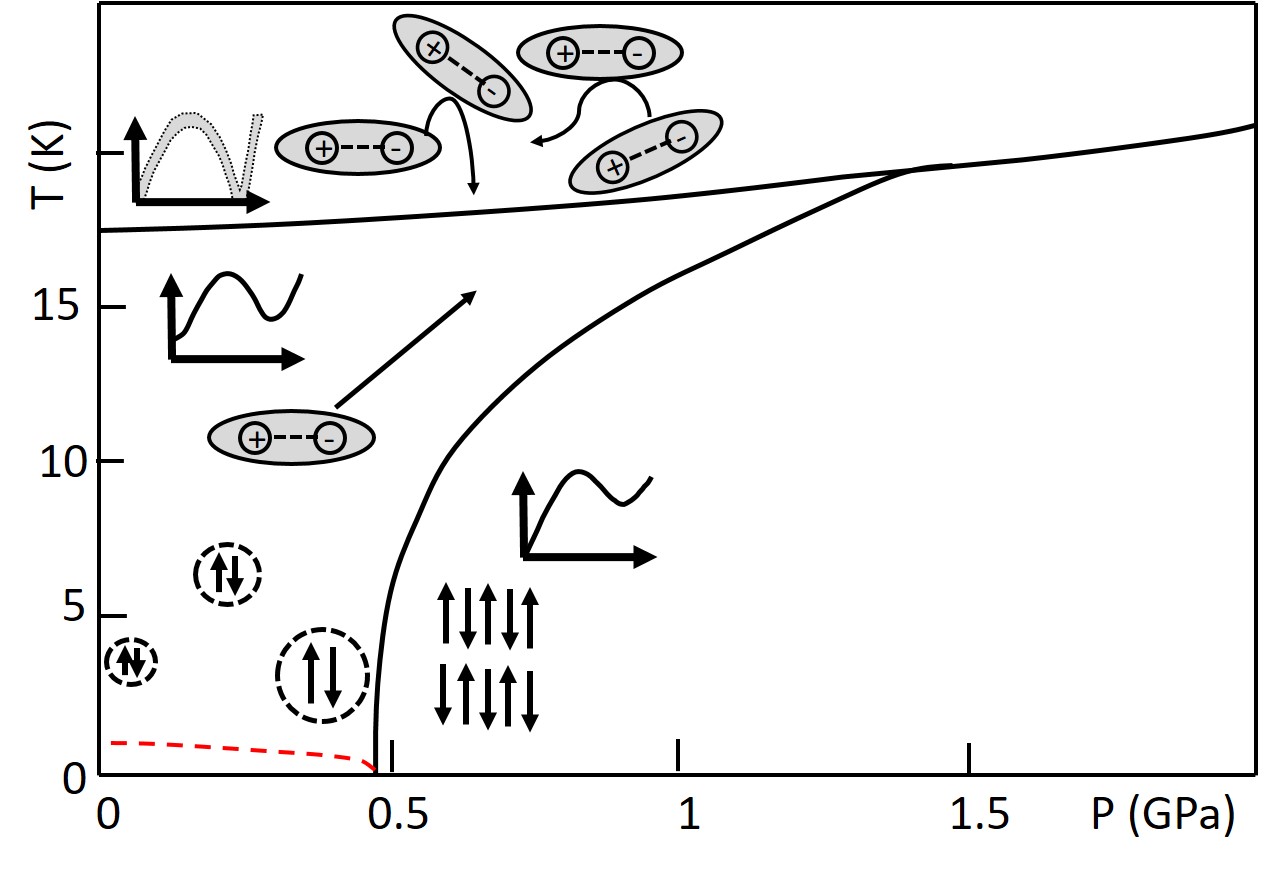}
	\end{center}
	\caption{A cartoon depiction of the excitations in the various parts of the temperature-pressure phase diagram of URu$_2$Si$_2$. For simplicity, we show the excitations as bound electron-hole pairs (excitons). In the paramagnetic phase (top part of diagram), the excitations are strongly damped and even over-damped (gapless) at $Q_0$ and $Q_1$ due to the sheer number present. This over-damping leads to degenerate regions in the dispersion, which in turn sets the stage for Kondo shielding. On cooling at low pressure, the number of excitations drops below a critical level and the excitations become under-damped (propagating). The concomitant appearance of a gap impedes Kondo screening and the resistivity jumps. The excitations themselves are propagating, explaining the sharp increase in thermal conductivity. Cooling further results in a superconducting phase (dotted line) of an unresolved nature. Increasing pressure in the HO-phase lowers the $Q_0$ gap (see text), resulting in larger droplets of the AF-phase. Once a critical pressure is reached, the $Q_0$ gap disappears \cite{bourdarot10} in a first-order transition and a long-range AF-phase is reached.}
	\label{phasediagram}
\end{figure}

Applying hydrostatic pressure at constant temperature to  URu$_2$Si$_2$ results in minor changes for moderate pressure, but results in a first-order transition at higher pressure. \cite{villaume} The ordered moment associated with the $Q_0$ Bragg peak increases with increased pressure, until a true long-range ordered AF state is reached above a critical pressure. We sketch this in Fig. \ref{phasediagram}. The AF-state and the HO-state are mutually exclusive. The HO-transition temperature increases slightly with increased pressure, until when the pressure is so high that the system transitions directly into the AF-state from the paramagnetic state upon lowering the temperature. \cite{villaume} Lastly, neutron scattering experiments in the HO-state  have shown that the $Q_1$ gap increases with pressure, whereas the $Q_0$ gap decreases with pressure. \cite{bourdarot10,williams} We next interpret these findings in terms of our thermally driven scenario.\\

The size of the gaps at $Q_0$ and $Q_1$ is determined by the RKKY-interaction mechanism $J$. Broholm {\it et al.} \cite{broholm} have shown that $J$ for the nearest neighbor (nn), next nearest neighbor (nnn), next-next nearest neighbor (nnnn) interactions are all negative. The (100) gap can then be attributed to the fact that each uranium ion has eight nnn along the body diagonal that add up in anti-phase, two nn along the a-direction and two nn along the b-direction (adding up in phase), and four nnnn along the (110)-direction (adding up in phase). According to Broholm {\it et al.} the nnn-interaction is the strongest so that $J_q$ for (100) is positive, and the dispersion $\omega_q$ reaches a minimum since $\omega_q=\sqrt{\Delta^2-4\alpha\Delta J_q}$ (with $\Delta$ the bare singlet-singlet gap, and $\alpha$ the transition element between the two singlet states \cite{broholm}). Increasing pressure increases the orbital overlap (and thereby $J$), resulting in a lowering the (100)-excitation gap.\\

The pressure evolution of $J_q$ for $q=Q_1$ is not as easily visualized due to the incommensurate nature, but apparently $J_q$ decreases slightly, resulting is a slightly larger gap $\omega_q$ at $q=Q_1$. This latter gap increase is the reason behind the slight increase in $T_{HO}$ with pressure. The number of excitations depends on the size of both gaps, but since the $Q_1$ gap has a fourfold weight in the Brillouin zone compared to the $Q_0$ gap, the critical number of excitons is determined by the $Q_1$ gap. As such, an increase in $Q_1$ gap leads to an increase in $T_{HO}$.\\

We note that the increase in $J$ at $Q_0$ with pressure (and concomitant decrease of the $Q_0$-gap) naturally drives the system to long-range AF-order, even when the excitations are between singlet states. Should small local moments develop on the U-ions, then having these moments probed in anti-phase (as $Q_0$ probing does) induces AF-order. Since this order lowers the energy of the system, URu$_2$Si$_2$ becomes unstable against moment formation with increased pressure.\\

The decrease in the $Q_0$ gap with increased pressure heralds the transition from the HO-state to the AF-state and the demise of the low-temperature superconducting state, as noted by various authors. \cite{bourdarot10,mydosh11} The transition appears to be first order, with the $Q_0$ gap only decreasing to about 1 meV before it closes in the AF-phase. This was uncovered in neutron scattering experiments \cite{villaume} at fixed pressure where cooling sees a transition into the HO-phase, followed by a transition into the AF-phase. With the disappearance of the $Q_0$ gap between the two singlet states, superconductivity disappears alongside as superconductivity requires a finite size gap to exist between the electronic ground state and first excited state.\\

In quantum critical systems, both hydrostatic pressure and chemical pressure can be used virtually interchangeably \cite{sullow} to drive the system between ordered and disordered phases. It appears that in URu$_2$Si$_2$ the situation is more complicated as chemical substitution results in both chemical pressure as well as in changes in hybridization. For instance, iso-electronic substitution of Fe on the Ru sites mimics the effects of hydrostatic pressure for low Fe concentrations \cite{das}, but the transition becomes more smeared out with increased Fe-concentration \cite{ran} and deviates at higher concentrations due to changes in hybridization \cite{wilson}. Amorese {\it et al.} \cite{severing} have shown by means of Xray scattering studies on fully substituted samples (UFe$_2$Si$_2$, UPd$_2$Si$_2$, and UNi$_2$Si$_2$) that while all systems are well described by a 5f2 state where the two lowest lying multiplets are singlets, 5f3 configurations get mixed in, with most mixing taking place in UFe$_2$Si$_2$ and the least amount in UPd$_2$Si$_2$ \cite{severing}. The authors linked the high degree of mixing in UFe$_2$Si$_2$ with the absence of a hidden order transition. This viewpoint is entirely consistent with our scenario of what drives the transition.

\subsection{Speculation and future experiments}

A potentially revealing experiment is using a heat-pulse method to measure the thermal conductivity (if at all possible in a metal). This method revealed the existence of second sound \cite{glydebook} in the superfluid phase. For reference, second sound refers to density oscillations in the gas of quasi-particles (as opposed to density oscillations in the gas of helium atoms). Should the excitations in URu$_2$Si$_2$ become very long lived, as they appear to do, then such second-sound oscillations could very well be present in URu$_2$Si$_2$. Should such oscillations be observed, then this firmly establishes the order parameter as the equivalent of the superfluid density.\\

A second experiment of potential interest was already proposed by Amorese {\it et al.} \cite{severing}: applying hydrostatic pressure to the Fermi-liquid UFe$_2$Si$_2$ compound should drive the system towards the hidden order phase as applying pressure favors the smaller ionic radius of the 5f2 configuration at the expense the 5f3 one. Should the hidden order transition emerge upon applying pressure but at a lower temperature, then we would expect to see large effects in the resistivity and thermal transport, but only a very modest jump in the specific heat as Kondo shielding will have removed a larger fraction of the entropy at lower temperature.\\

Zero-point-motion responsible for the spread in Kondo temperatures could very well play a role in the observed steady increase in AF-moment with increasing pressure. While we correctly think of pressure as being uniform throughout the sample, we tend to also think of the effects induced by pressure as being uniform throughout the sample. However, when it comes to the interatomic separations, this is incorrect. Spontaneous zero-point motion leads to 1-2\% changes in these separations \cite{RuU}, a magnitude comparable to the average pressure induced changes. Since electrons move much faster than the ions, to the electrons these spontaneous ionic displacements appear to be static displacements, as if locally the pressure were not uniform. In local regions corresponding to decreased inter-atomic separations, the system is effectively in the high-pressure AF-phase and locally there will be an ordered staggered moment. Of course, with increased overall pressure, we expect to find more and more such regions as the average inter-atomic separations are already smaller, resulting in the overall observed increase of AF-moment with increased pressure. \\

Thus, the observed AF-droplets could simply be the result of quantum fluctuations, with the observed AF-moment that of small regions in the AF-phase (the standard interpretation). The subtlety in viewing these droplets as being caused by zero-point-motion is that the droplets are an intrinsic, unavoidable part of the HO-phase where small regions are driven to the AF-phase because of unavoidable quantum fluctuations, with the HO-AF transition related to these droplets having grown so much in number that they form a lattice spanning network. Note that the only essential assumption in the above is that the electronic timescales are much faster than the ionic timescales, so that to electrons the ionic displacements are static.\\

\subsection{Summary}

In summary, we have depicted the similarities between the hidden order transition in URu$_2$Si$_2$ and the superfluid transition in $^4$He. In both systems a gap materializes at the transition temperature detected by inelastic neutron scattering at distinct momentum values. These gapped elementary excitations (EEs) determine the character of the phase transitions and order parameters (OP). Unfortunately, for URu$_2$Si$_2$ the exact nature of the EEs and OP have not yet been clearly established. While the dispersion of  URu$_2$Si$_2$ is a continuous one throughout the Brillouin zone, almost certainly representing excitations out of a singlet ground state, the different pressure dependence of the $Q_0$ and $Q_1$ EEs indicates \cite{butch} that they play a slightly different role in the HO phase as they would result in different ground states should either gap close at low temperature. Therefore, based on our comparisons with $^4$He we propose that the two singlet-singlet EEs of URu$_2$Si$_2$ at $Q_0$ and $Q_1$ take on the character of antiferromagnetic fluctuations at the (1,0,0) commensurate wave-vector (brought about by an instability against moment formation and enhanced by hydrostatic pressure) and of charge-density-wave oscillations at the incommensurate (0.6,0,0) wave-vector (enhanced by putative negative pressure). Of course, in the absence of any (positive or negative) pressure, neither minimum has succeeded in driving the system to a long-range ordered ground state.\\

Loosely speaking, the gap opens itself up and does not require any change in underlying symmetry to emerge. Overall, we have a fundamental change in the nature of the phases that is precipitated by the number of excitations thermally excited.  As such the OP governing the HO transition bears similarities to $^4$He in that both are of the continuous, second-order kind that are reflected in the entropy and transport properties. The opening up of the gap that drives the transition sets the stage for the low-temperature ground state to materialize. In the case of helium it is the emergence of a Bose-Einstein condensate, in the case of URu$_2$Si$_2$ it could be any of the proposed configurations such as, for instance, a chirality density wave \cite{chiral}. However, in the absence of the need to explain the specific heat, resistivity, and thermal conductivity jumps (as the unavoidable opening up of a gap explains all of this), we should also keep an open mind to the possibility that the hidden order state is nothing more that a gas of singlet excitations in a nearly compensated metal, which does not become puzzling again until superconductivity materializes (potentially mediated by the AF-fluctuations). Independent of the exact details of the ground state, our scenario proposes that it is the opening up of the gap that sets the stage for any order to emerge, it is not the low-temperature ground state configuration that drives the opening of a gap between the ground state and the excited states.\\

In a simplified nutshell, the HO transition is effectively comparable to the harmonic oscillator: when the damping is increased there is a reduction in the oscillation frequency even to the point where the oscillator can become over-damped, i.e., no EEs. In URu$_2$Si$_2$ and $^4$He the damping is caused by the number of excitations (bosons) present and this number increases rapidly once the gap closes due to the increased damping. When this happens, the excitations are no longer elementary excitations, but rather short-lived excitations that cannot travel far before decaying. We summarize this scenario in Fig. \ref{phasediagram}. The temperature dependence of the entropy and transport (resistivity and thermal conductivity) properties follow naturally from the number of excitations present combined with whether they propagate or not. Independent of the exact nature of the EEs, a thermally activated gap explains how URu$_2$Si$_2$ can exhibit a $\lambda$-anomaly in the specific heat without any obvious signs of an order parameter being present.\\

Within our sketched scenario for the hidden order transition in URu$_2$Si$_2$, it is natural to ask whether URu$_2$Si$_2$ is a unique system, or whether more such systems are waiting to be discovered. What is required in our scenario are two energy levels that are close enough in energy that a sufficient number of quasi-particles can be excited by raising the temperature (without affecting other parts of the system) for the levels to become degenerate. For URu$_2$Si$_2$ and helium, this number is reached for $\Delta/(k_BT_{transition}) \sim 3-4$, with $\Delta$ the gap energy of the driving minimum. We expect that any system where the temperature can be raised to a few times the gap energy without mixing in other degrees of freedom will exhibit a marked change in thermal conductivity. In order to see corresponding changes in the resistivity of a metal we would also need that the lowest lying levels are singlets in order to affect the Kondo screening mechanism. In addition to the singlet states requirement, the value of the average Kondo-shielding temperature also plays a role. Not only does the average $T_K$ determine how much entropy is left to be shed at the transition, since both $T_K$ and the RKKY-ordering temperature are determined \cite{doniach} by the same orbital overlap mechanism, the window for optimal shielding temperature appears to be quite narrow. A high shielding temperature would imply a large energy gap between the singlet states, making it very unlikely that the temperature can be raised enough without other degrees of freedom being mixed in (breaking the isolation of the low-lying singlets), whereas a low shielding temperature would favor moment formation and long-range magnetic order as exemplified in the Doniach \cite{doniach} phase diagram. As such, URu$_2$Si$_2$ might well be a unique system, but experiments under pressure on UFe$_2$Si$_2$ as proposed by Amoresi {\it et al.} \cite{severing} might well reveal another hidden-order compound.

\end{document}